\documentclass[]{aastex631}
\usepackage{amsmath}

\usepackage[utf8]{inputenc}
\DeclareUnicodeCharacter{02BC}{'}  

\usepackage{threeparttable}
\usepackage{footnote}
\usepackage{tabularx}
\usepackage[figuresright]{rotating}
\usepackage{appendix}

\begin{document}

\shortauthors{Tao-zhi Yang et al.}
\shorttitle{KIC 5623923: A Faint Eclipsing Binary Consisting of $\delta$ Scuti pulsations}
\title{KIC 5623923: A Faint Eclipsing Binary Consisting of $\delta$ Scuti pulsations }

\author[0000-0002-1859-4949]{Tao-Zhi Yang}
\affiliation{Ministry of Education Key Laboratory for Nonequilibrium Synthesis and Modulation of Condensed Matter, School of Physics, Xi'an Jiaotong University, 710049 Xi'an, People's Republic of China}

\author[0000-0001-6693-586X]{Zhao-Yu Zuo}
\affiliation{Ministry of Education Key Laboratory for Nonequilibrium Synthesis and Modulation of Condensed Matter, School of Physics, Xi'an Jiaotong University, 710049 Xi'an, People's Republic of China} 
\affiliation{Institut f{\"u}r Astronomie und Astrophysik, Kepler Center for Astro and Particle Physics, Eberhard Karls, Universit{\"a}t, Sand 1, D-72076 T{\"u}bingen, Germany}
\email{zuozyu@xjtu.edu.cn}

\author{Shi-ping Guo}
\affiliation{Ministry of Education Key Laboratory for Nonequilibrium Synthesis and Modulation of Condensed Matter, School of Physics, Xi'an Jiaotong University, 710049 Xi'an, People's Republic of China}

\author{Xu Ding}
\affiliation{Yunnan Observatories, Chinese Academy of Sciences (CAS), P.O. Box 110, 650216 Kunming, People’s Republic of China}
\affiliation{Key Laboratory of the Structure and Evolution of Celestial Objects, Chinese Academy of Sciences, P. O. Box 110, 650216 Kunming, People’s Republic of China}
\affiliation{Center for Astronomical Mega-Science, Chinese Academy of Sciences, 20A Datun Road, Chaoyang District, Beijing, 100012, People’s Republic of China}

\author{Hao-zhi Wang}
\affiliation{Xinjiang Astronomical Observatory, Chinese Academy of Sciences, Urumqi, Xinjiang 830011, China}
\affiliation{School of Astronomy and Space Science, University of Chinese Academy of Sciences, Beijing 100049, Peopleʼs Republic of China}

\author{Shahidin Yaqup}
\affiliation{Xinjiang Astronomical Observatory, Chinese Academy of Sciences, Urumqi, Xinjiang 830011, China}

\author{Ali Esamdin}
\affiliation{Xinjiang Astronomical Observatory, Chinese Academy of Sciences, Urumqi, Xinjiang 830011, China}
\affiliation{School of Astronomy and Space Science, University of Chinese Academy of Sciences, Beijing 100049, Peopleʼs Republic of China}

\begin{abstract}

In this paper, we present a detailed analysis of the light variation of KIC 5623923 using high-precision time-series data from the $Kepler$ mission. The analysis reveals this target is an eclipsing binary system with $\delta$ Scuti type pulsations from the primary component, rather than from the secondary as previously reported. The frequency analysis of three short-cadence data reveals 41 significant frequencies, including the orbital frequency ($f_{orb}$ = 0.827198 d$^{-1}$) due to orbital motion from binary system and the pulsational frequencies. Most of the pulsational signal lies in the frequency range of 20 - 32 d$^{-1}$, with amplitude between 0.3 and 8.8 mmag, in which seven peaks are identified as `independent' modes. The strongest one ($f_{3}$ = 28.499399 d$^{-1}$) likely corresponds to a high-order radial mode. In other peaks ($f_{7}$, $f_{10}$, and $f_{18}$), several pairs of multiplet structures centered on them are found. The fitting of spectral energy distribution (SED) using the collected photometry measurement of multiple bands reveals the effective temperatures of the primary and secondary components as $8348^{+230}_{-225}$~K and $4753^{+237}_{-229}$~K, respectively, which place the primary star in the classical pulsating instability zone. The characteristic light curve morphology and short orbital period are consistent with a tidally locked system. Based on the characteristics of amplitude spectra of pulsating stars in close binaries, the analysis of the multiplet structures reveals that three independent frequencies (i.e. $f_{7}$, $f_{10}$, and $f_{18}$) correspond to non-radial modes with $l = 2$, while the associated sidelobes are produced by the orbital motion. We highlight the potential of this method in future studies of pulsating binary stars.

\end{abstract}


\section{Introduction}

Eclipsing binary systems (EBs) are important sources for determining the physical parameters of stars \citep{1980ARA&A..18..115P,2010A&ARv..18...67T}.  With high-precision photometric and spectroscopic observations, accurate stellar parameters, such as the mass, radius, and effective temperatures of components in EBs can be determined precisely within 1 percent, which are very helpful for testing and calibrating already existing stellar models of evolution \citep{2005MNRAS.363..529S,2008A&A...487.1095C,2012ocpd.conf...51S}. 
EBs containing pulsating components, also called pulsating EBs, are particularly valuable since fundamental stellar parameters (mass, radius, effective temperature) can be determined from two independent ways: EBs and pulsations. Pulsations, which reflect the brightness variations on the surface of a star, provide a unique opportunity to probe the internal properties and processes with asteroseismolgy \citep{2010aste.book.....A}. Furthermore, independently measured properties from EBs can also be used to improve asteroseismic modelling for pulsating stars, in which mode identifications are not straightforward.

The synergy between binary systems and the asteroseismology of pulsating stars has attracted great scientific interest over the past three decades. \cite{1990JAVSO..19...52S} presented the first list of pulsating stars in binary systems with 58 members. Since then, the number of pulsating stars in binary systems has steadily increased, encompassing not only $\delta$ Sct stars \citep{2000A&AS..144..469R,2001A&A...366..178R}, but also other types, such as solar-like oscillators and pulsating sdB/O stars. Thanks to the ultra-high-precision and continuous photometric observations, such as CoRoT \citep{2009A&A...506..411A}, $Kepler$ \citep{2010Sci...327..977B,2010ApJ...713L..79K}, and TESS \citep{2015JATIS...1a4003R}, the number of new binary and multiple systems with pulsating components is rapidly increasing. \cite{2010arXiv1002.2729Z} published a catalogue of 515 pulsating binary/multiple systems as of October 2014, among which more than half of them are pulsating EBs, making the number of known oscillating Algol-type EBs ('oEA', \citealt{2002ASPC..259...96M}) to be 96. Subsequently, \cite{2017MNRAS.465.1181L} presented a list with 199 binary systems that host at least one pulsating component of $\delta$ Sct-type. With Kepler data, \cite{2019A&A...630A.106G} identified 303 systems (163 new), including 149 $\delta$ Sct (100 new), 115 $\gamma$ Dor (69 new), 85 red giants (27 new), and 59 tidally excited oscillators (29 new). With TESS observations, some systematic searches also resulted in hundreds of pulsators of various types in EBs, including 54 new pulsating stars in EA-type binary systems \citep{2022ApJS..259...50S}, 242 pulsators (143 new) \citep{2022ApJS..263...34C}, and 78 $\beta$ Cep variables (59 new) \citep{2024ApJS..272...25E}.

Among them, EBs containing $\delta$ Sct-type oscillations have great scientific potential for stellar astrophysics. 
$\delta$ Sct stars, as a certain class of short-period pulsating stars, are situated at the intersection of low-mass stars with radiative cores and convective envelopes, and high-mass stars with convective cores and radiative envelopes. They are characterized by thin surface convective envelopes, with their pulsations primarily driven by the $\kappa$ mechanism operating in the He II ionization zone \citep{2017ampm.book.....B, 2018MNRAS.476.3169B}.
They are also considered as transition objects from high-amplitude radial pulsators, such as Cepheid variables, to the non-radial multiperiodic pulsators within the classical instability strip \citep{2000ASPC..210....3B}. Most $\delta$ Sct stars are multiperiodic variables with periods in the range of 0.02-0.25 days \citep{2011A&A...534A.125U,2014MNRAS.439.2078H}, and show low-order radial/non-radial pressure (p) modes 
\citep{1963ApJ...138..487C,2010aste.book.....A,2018MNRAS.476.3169B}. The masses of this type of stars range from 1.5 to 2.5 M$\odot$, and they have effective temperatures in the range between 6000 K and 9000 K, covering the spectral types from A0 to F5. In the Hertzsprung–Russell (H-R) diagram, they are located at the intersection of the main sequence and the classical Cepheid instability strip \citep{2001A&A...366..178R}. Thus, these pulsations are usually used to probe the envelope of a star with asteroseismology \citep{1999A&A...351..582H,2021FrASS...8...55G}. 

The $\delta$ Sct components in binaries share similar features with their single counterparts, but the pulsations in oEA may be affected by the companion due to mass transfer and gravitational interaction \citep{2006MNRAS.366.1289S}. Based on an analysis of over 100 pulsating eclipsing binaries, \cite{2015ASPC..496..195L,2016arXiv160608638L} identified a strong correlation between $P_{orb}$ and $P_{pul}$, with an orbital period threshold of about 13 days, below which the pulsations appear to be significantly influenced by binarity. \citet{2013ApJ...777...77Z} investigated the relationship between orbital and pulsation periods from a theoretical perspective and found that pulsations in EBs depend on multiple factors, including the orbital period, the mass ratio, and the filling factor of the primary pulsating component. In addition, tidally excited modes have been identified in binaries with eccentric orbits, where the excited frequencies typically occur at integer multiples of the orbital frequency \citep{2011ApJS..197....4W,2013MNRAS.434..925H}. Some pulsating EBs at special evolutionary stages have also been reported. For example, KIC 8840638 is a recently discovered pulsating EB exhibiting $\delta$ Scuti–type oscillations and is believed to have undergone a mass ratio reversal, where the more massive star is the gainer and the less massive one is the donor \citep{2024ApJ...975..171Y}. However, how mass transfer affects the internal structure and pulsation behavior of stars remains an open question that requires further investigation.

KIC 5623923 ($\alpha_{2000}$=$19^{h}$$32^{m}$$01^{s}$.5, $\delta_{2000}$=+$40^{\circ}$$51^{'}$$16^{''}$.8, 2MASS: J19320153+4051166) was classified as a $\delta$ Scuti star in a contact binary by \citet{2014MNRAS.437..132R}, based on the typical light curve shape of binary stars and the detection of pulsations with a period of approximately 50 minutes. As the pulsations are not clearly visible during secondary eclipses, the secondary star (i.e., the less luminous component) is considered the likely source of the variability \citep{2014MNRAS.437..132R}. {Furthermore, \cite{2019A&A...630A.106G} reported KIC 5623923 as a pulsating binary system exhibiting $\delta$ Scuti pulsations and determined its configuration to be semi-detached}. Table \ref{tab:basic_parmeters} lists some basic parameters of KIC 5623923 collected from the survey and $Kepler$ Input Catalog (KIC; \citealt{2011AJ....142..112B}). In this paper, we further investigate the light variability and astrophysical properties of the system using high-precision $Kepler$ photometric time-series data.

\begin{table*}
\begin{center}
\caption{Basic parameters of KIC 5623923.\label{tab:basic_parmeters}}
\begin{tabular}{lcc} \hline \hline
\noalign{\smallskip}
Parameters & KIC 5623923 &  References \\
\hline
\noalign{\smallskip}

 $P_{orb}$      &  1.21 day &   a  \\
 $P_{pul}$      &  50 min   &   a  \\
 $T_{KIC}$      &  8300 $\pm$ 250 K  &   b  \\
 $T_{GTC}$      &  7970 $\pm$ 110 K  &   a  \\
 $\log g$       &  4.25  $\pm$ 0.25 dex & c  \\
\noalign{\smallskip}            
\hline   
\end{tabular}
\end{center}

\tablecomments{(a) \citealt{2014MNRAS.437..132R}; (b) $Kepler$ Input Catalog \citealt{2011AJ....142..112B}; (c) https://gea.esac.esa.int/archive/. Note that the KIC temperatures are systematically smaller than other temperature determinations \citep{2014ApJS..211....2H}, while $T_{GTC}$ is determined from spectroscopy (INT + GTC spectra obtained at Obs. Roque de los Muchachos, La Palma) \citep{2014MNRAS.437..132R}.}
\end{table*}

\section{Observations and Data Reduction}

\begin{figure*}
  \centering
  \includegraphics[width=0.95\textwidth,trim=30 245 25 245,clip]{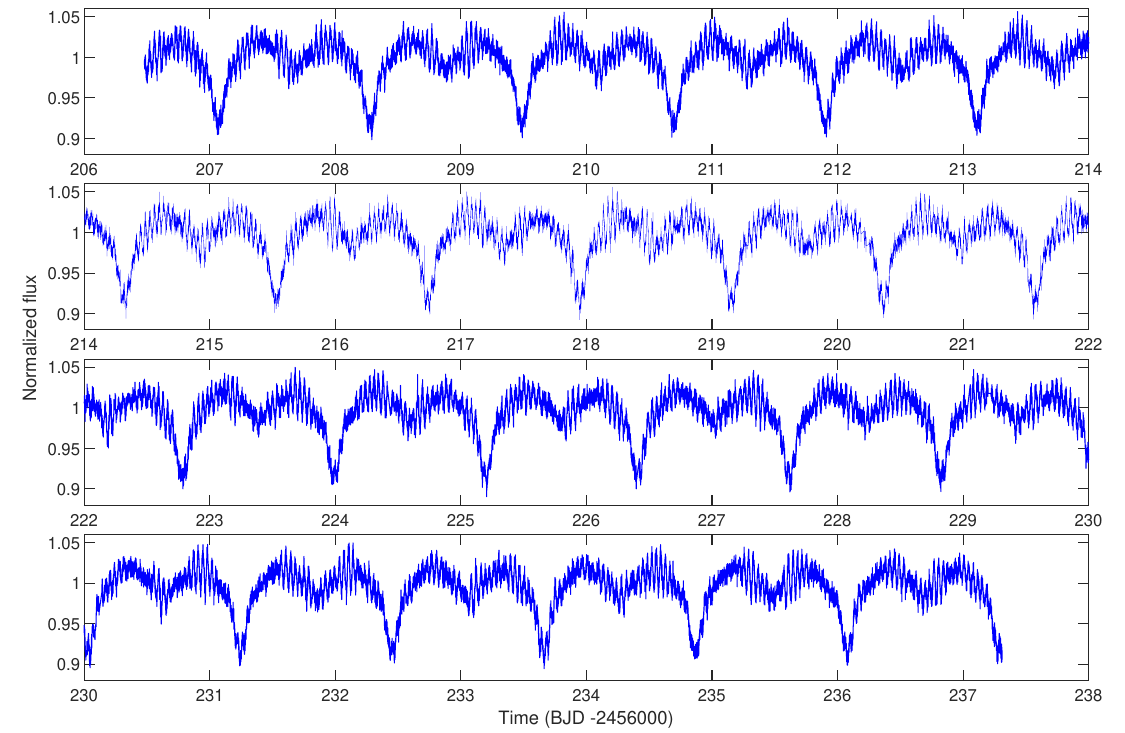}
\caption{The light curve of KIC 5623923 in full SC Quarter 14.1, clearly showing the typical light variations of an eclipsing binary system with superposition of pulsations.}
\label{fig:Fig1_light_curve}
\end{figure*}

KIC 5623923 was observed by $Kepler$ space telescope from BJD 2456107.13 to 2456414.59, which spans 307.46 days. There are 4 quarters (Q14.1, Q15.1, Q16.1, and Q17.1) of short cadence (SC, 59 s integration time ) data, containing 119898 points in total, as well as 4 quarters (Q14, Q15, Q16, and Q17) of long cadence (LC, 29.5 minute integration time) data. Considering the high-sampling rate of SC data, we only use the SC data in this work. The SC photometric flux data of KIC 5623923 is available in $Kepler$ Asteroseismic Science Operations Center (KASOC) database\footnote{KASOC database: {http://kasoc.phys.au.dk}}\citep{2010AN....331..966K} with two types: the first is labeled as 'raw' data which was produced by the NASA $Kepler$ Science pipeline, and the second is the flux data corrected by KASOC Working Group 4 (WG$\#$4: $\delta$ Scuti targets). We downloaded all the SC time-series data of this target and used the corrected data. For each quarter, several obvious outliers were first removed with 4$\sigma$ clipping and possible linear or slow trend were also corrected {using a linear or low-order (2nd or 3rd) polynomial fitting}. Then all the data were stitched to a total light curve with 119898 data points. Figure \ref{fig:Fig1_light_curve} shows the light curve of KIC 5623923 in SC Quarter 14.1. It is clear that its light curve shows a superposition of pulsations and an additional cyclic variation, which presents the typical light variations of an eclipsing binary system.

\section{Pulsational Characteristics}

To investigate the pulsating behavior of KIC 5623923, we performed Fourier analysis on the SC time-series data using the software PERIOD04 \citep{2005CoAst.146...53L}, which is especially dedicated to the statistical analysis of astronomical time series containing gaps. To detect more significant frequencies in SC data, we chose a frequency range of 0 $<$ $\nu$ $<$ 80 d$^{-1}$, a little wider than the typical pulsation frequency range of $\delta$ Sct stars.

{Significant frequencies were extracted iteratively. In each iteration, the highest peak in the amplitude spectrum was selected, and a multi-frequency least-squares fit was performed using the formula:} $m = m_{0} + \Sigma\mathnormal{A}_{i}sin(2\pi(\mathnormal{f}_{i}\mathnormal{t} + \phi_{i}))$ ($m_{0}$ is the zero-point, $A_{i}$ is the amplitude, $f_{i}$ is the frequency, and $\phi_{i}$ is the corresponding phase) for the light curve, resulting in the solution of all frequencies. A constructed light curve using the above solutions was subtracted from the data, and the residual was obtained to search for the next significant frequency. Then, the above steps were repeated until there was no significant peak in the frequency spectrum. The criterion (S/N $>$ 4.0) suggested by \citet{1993A&A...271..482B} was adopted to judge the significant peaks. The uncertainties of the frequencies were calculated following the method of \citet{2008A&A...481..571K}.

\begin{figure*}
\begin{center}
  \includegraphics[width=0.95\textwidth,trim=30 245 25 245,clip]{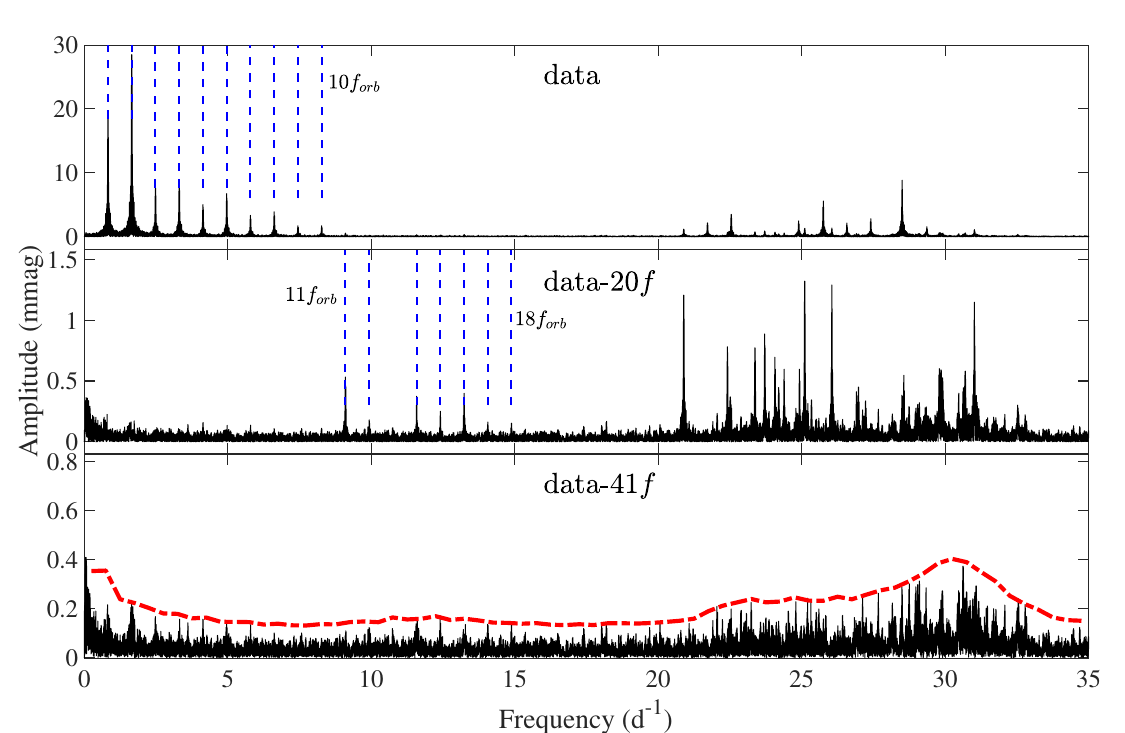}
  \caption{Top panel: the amplitude spectrum of SC data for KIC 5623923 up to 35 d$^{-1}$, the blue dashed lines refer to the positions of the orbital frequency $f_{orb}$ and its harmonics up to 10$f_{orb}$. Middle panel: the amplitude spectrum after prewhitening of 20 strongest peaks, the blue dashed lines refer to harmonics of the orbital frequency up to 18$f_{orb}$. Bottom panel: the residual spectrum after extraction of all significant frequencies, to an amplitude limit of about 0.1 mmag. The red dot-dashed curve refers to the detection limit of S/N = 4.0. No peak is statistically significant in the residual.}
    \label{fig:SC_spectra}
\end{center}
\end{figure*}

\begin{figure} 
  \centering
  \includegraphics[width=0.95\textwidth,trim=30 245 25 245,clip]{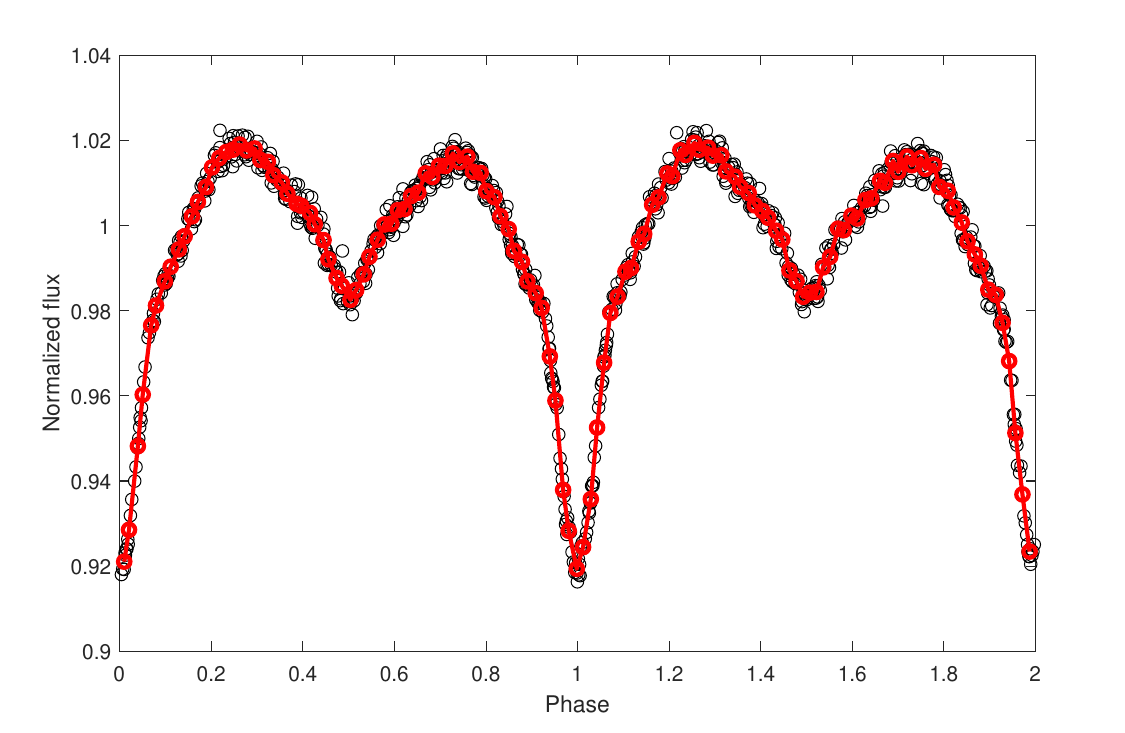}
   \caption{{Phase-folded light curve of KIC 5623923 on the orbital frequency ($f_{1}$), with pulsations removed. The red circles represent the data binned into 100 phase intervals.}}
   \label{fig:SC_phase}
\end{figure}

A total of 41 significant frequencies were detected and listed in Table \ref{tab:Frequency-SC}. Figure \ref{fig:SC_spectra} shows the Fourier amplitude spectra and the prewhitening process for the light curve of KIC 5623923. Note that almost all detected frequencies are less than 32.0 d$^{-1}$, except for a harmonic of the dominant frequency. All these frequencies are divided into two groups: one lies in the low frequency region 0 - 15 d$^{-1}$ and the other in 20 - 35 d$^{-1}$. In the lower frequency region, the frequency $f_{1}$ (=0.827198 d$^{-1}$) can easily be identified as the orbital frequency, due to the orbital motion of the eclipsing binary system. Figure \ref{fig:SC_phase} shows the phase diagram of the SC data, which includes 100 time-bins. To clearly show light variations due to binary orbital motion, the light curve is folded with the detected low frequency of $f_{1}$. From this figure, the orbital phase diagram of KIC 5623923 is clearly shown, so the lowest frequency $f_{1}$ was marked as the orbital frequency (denoted `$f_{orb}$') in Table \ref{tab:Frequency-SC}. Moreover, a series of harmonic (see blue dashed lines in the top and middle panels of Fig. \ref{fig:SC_spectra}) of orbital frequency were also detected up to 18$f_{orb}$, due to the non-sinusoidal shape of the light curve.

\startlongtable
\begin{deluxetable*}{crcrcc}
\tabletypesize{\small}
\tablewidth{0pc}
\tablecaption{All frequencies detected in SC data.\label{tab:Frequency-SC}}
\tablehead{
\colhead{$f_{i}$}   &
\colhead{Frequency (d$^{-1}$)}  &
\colhead{Amplitude (mmag)}      &
\colhead{S/N}            &
\colhead{Identification} &
}
\startdata
 1 &  0.827198 $\pm$ 0.000007 &     18.68 $\pm$ 0.19 &  170.0 &  $f_{orb}$  &   \\
 2 &  1.654610 $\pm$ 0.000003 &     28.66 $\pm$ 0.12 &  396.7 &  2$f_{orb}$    &   \\
 3 & 28.499399 $\pm$ 0.000015 &      8.82 $\pm$ 0.20 &  76.5  &  independent   &   \\
 4 &  3.309224 $\pm$ 0.000006 &      8.70 $\pm$ 0.08 &  179.1 &  4$f_{orb}$ &   \\
 5 &  2.481911 $\pm$ 0.000009 &      7.79 $\pm$ 0.11 &  125.2 &  3$f_{orb}$  &   \\
 6 &  4.963835 $\pm$ 0.000018 &      6.73 $\pm$ 0.18 &  62.6  &  6$f_{orb}$  &   \\
 7 & 25.753836 $\pm$ 0.000036 &      5.60 $\pm$ 0.32 &  30.4  &  independent  &   \\
 8 &  4.136520 $\pm$ 0.000010 &      5.08 $\pm$ 0.08 &  109.2 &  5$f_{orb}$  &   \\
 9 &  6.618418 $\pm$ 0.000039 &      3.88 $\pm$ 0.23 &  28.6  &  8$f_{orb}$  &   \\
10 & 22.545359 $\pm$ 0.000037 &      3.51 $\pm$ 0.20 &  29.6  &  independent  &  \\    
11 &  5.791157 $\pm$ 0.000062 &      3.30 $\pm$ 0.32 &  17.8  &   7$f_{orb}$  &  \\
12 & 27.408428 $\pm$ 0.000087 &      2.86 $\pm$ 0.39 &  12.7  &   f7+2$f_{orb}$  & \\
13 & 24.895582 $\pm$ 0.000038 &      2.44 $\pm$ 0.14 &  28.9  &   independent & \\
14 & 21.717935 $\pm$ 0.000117 &      2.16 $\pm$ 0.39 &   9.5  &   f10-$f_{orb}$  & \\
15 & 26.576125 $\pm$ 0.000096 &      2.15 $\pm$ 0.32 &  11.5  &   f7+$f_{orb}$  &  \\
16 &  7.445680 $\pm$ 0.000087 &      1.69 $\pm$ 0.23 &  12.8  &   9$f_{orb}$  &  \\
17 &  8.273104 $\pm$ 0.000062 &      1.67 $\pm$ 0.16 &  17.8  &   10$f_{orb}$   &  \\
18 & 29.363803 $\pm$ 0.000087 &      1.56 $\pm$ 0.21 &  12.7  &  independent  & \\
19 & 25.107480 $\pm$ 0.000038 &      1.32 $\pm$ 0.08 &  28.9  &  independent  & \\
20 & 26.051505 $\pm$ 0.000117 &      1.29 $\pm$ 0.23 &   9.5  &  f18-4$f_{orb}$  & \\
21 & 20.890673 $\pm$ 0.000096 &      1.21 $\pm$ 0.18 &  11.5  &  f10-2$f_{orb}$  &  \\
22 & 31.018652 $\pm$ 0.000087 &      1.15 $\pm$ 0.15 &  12.8  &  f18+2$f_{orb}$  &  \\
23 & 23.710849 $\pm$ 0.000062 &      0.88 $\pm$ 0.08 &  17.8  &  independent   &  \\
24 & 22.413770 $\pm$ 0.000087 &      0.78 $\pm$ 0.11 &  12.7  &  f13-3$f_{orb}$  & \\
25 & 23.372756 $\pm$ 0.000038 &      0.76 $\pm$ 0.05 &  28.9  &  f10+$f_{orb}$  & \\
26 & 24.068293 $\pm$ 0.000117 &      0.69 $\pm$ 0.12 &   9.5  &  f13-$f_{orb}$  & \\
27 & 24.921574 $\pm$ 0.000096 &      0.64 $\pm$ 0.10 &  11.5  &  f7-$f_{orb}$  &  \\
28 & 24.388823 $\pm$ 0.000117 &      0.61 $\pm$ 0.11 &  9.5   &   - &  \\
29 & 24.199762 $\pm$ 0.000252 &      0.60 $\pm$ 0.23 &  4.4   &  f10+2$f_{orb}$  &  \\
30 & 30.699680 $\pm$ 0.000236 &      0.58 $\pm$ 0.21 &  4.7   &   - &  \\
31 & 28.562605 $\pm$ 0.000171 &      0.56 $\pm$ 0.15 &  6.5   &   - &  \\
32 &  9.100338 $\pm$ 0.000074 &      0.53 $\pm$ 0.06 &  14.9  &  11$f_{orb}$  &  \\
33 & 56.998787 $\pm$ 0.000074 &      0.51 $\pm$ 0.06 &  14.9  &  2f3  &  \\
34 & 26.907656 $\pm$ 0.000163 &      0.41 $\pm$ 0.10 &  6.8   &   - &  \\
35 & 13.236646 $\pm$ 0.000087 &      0.40 $\pm$ 0.05 &  12.7  &  16$f_{orb}$  &  \\
36 & 11.582286 $\pm$ 0.000098 &      0.38 $\pm$ 0.06 &  11.3  &  14$f_{orb}$  &  \\
37 & 25.342607 $\pm$ 0.000231 &      0.32 $\pm$ 0.11 &  4.8   &   - &  \\
38 & 12.409547 $\pm$ 0.000135 &      0.25 $\pm$ 0.05 &  8.2   &  15$f_{orb}$  &  \\
39 &  9.927762 $\pm$ 0.000213 &      0.18 $\pm$ 0.06 &  5.2   &  12$f_{orb}$  &  \\
40 & 14.064137 $\pm$ 0.000222 &      0.17 $\pm$ 0.06 &  5.0   &  17$f_{orb}$  &  \\
41 & 14.880700 $\pm$ 0.000213 &      0.15 $\pm$ 0.05 &  5.2   &  18$f_{orb}$  &  \\
    \enddata
\end{deluxetable*}

We examined the frequencies in the region of 20-35 d$^{-1}$ and found 7 strong frequencies (i.e. $f_{3}$, $f_{7}$, $f_{10}$, $f_{13}$, $f_{18}$, $f_{19}$, and $f_{23}$) can be considered independent frequencies, since they are neither combinations nor harmonics of other frequencies. These seven frequencies are marked with `independent' in the last column of Table \ref{tab:Frequency-SC}. Other frequencies can be identified as combinations with a simple form ($f_{}$ $\pm$ $\sigma$) = $m$($f_{1}$ $\pm$ $\sigma _1$) $\pm$ $n$($f_{2}$ $\pm$ $\sigma _2$) using the resolution (1.5/$\Delta$T = 0.0049 d$^{-1}$), where $m$ and $n$ are small integers, $f_{1}$ and $f_{2}$ are the stronger frequencies, and we denoted their identifications in the last column of Table \ref{tab:Frequency-SC}. 

\begin{figure*}
\begin{center}
  \includegraphics[width=0.95\textwidth,trim=30 245 25 245,clip]{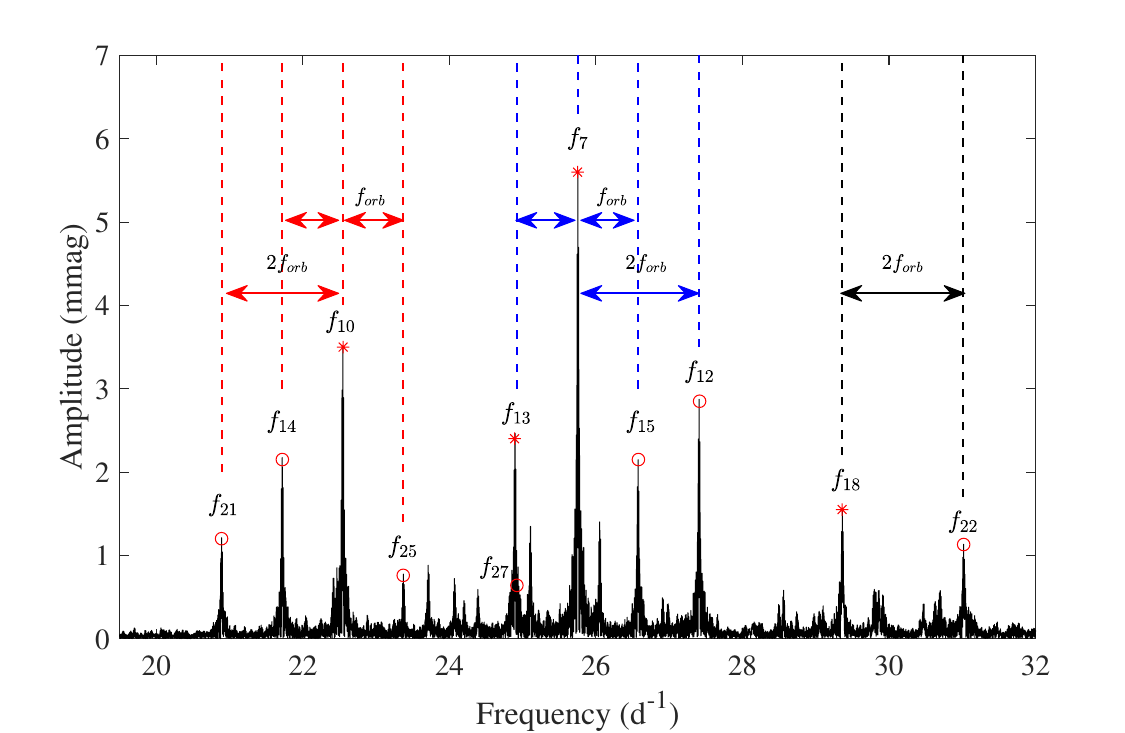}
  \caption{Amplitude spectrum of SC data for KIC 5623923 in frequency region of 20 - 32 d$^{-1}$, clearly showing the multiplet structures around $f_{7}$, $f_{10}$, and $f_{18}$. Asterisk ($\ast$) refers to positions of the independent frequencies and circle ($\circ$) refers to the combinations with $f_{orb}$. }
    \label{fig:SC_spectra2}
\end{center}
\end{figure*}

Among these frequencies, there are two groups: ($f_{7}$, $f_{12}$, $f_{13}$, and $f_{15}$) and ($f_{10}$, $f_{14}$, $f_{21}$, and $f_{25}$), in which each group forms a comb-like structure. On closer inspection, we found that they have equal-spacing with the value of $f_{orb}$, as shown in Figure \ref{fig:SC_spectra2}. Since the stronger peaks, $f_{7}$ and $f_{10}$, are independent pulsation frequencies, the other frequencies of each group are considered as the sidelobes which are split from the central peaks (i.e. $f_{7}$ and $f_{10}$) by exactly the orbital frequency (i.e. $f_{orb}$), which has been discussed in detail by \cite{2012MNRAS.422..738S}. The frequency, $f_{18}$, was found to have only one split peak, $f_{22}$. The possible reason is that its amplitude is too weak. 

The frequency $f_{3}$, the strongest peak among the independent frequencies, is most interesting, as there is no comb-like structure similar to that found around $f_{7}$ and $f_{10}$. Moreover, its harmonic (i.e. $f_{33}$) has also been detected, while $f_{7}$ and $f_{10}$ do not. Therefore, $f_{3}$ may belong to different pulsation modes, compared to frequencies such as $f_{7}$ and $f_{10}$. The remaining five weaker frequencies (namely, $f_{28}$, $f_{30}$, $f_{31}$, $f_{34}$, and $f_{37}$) {could not be reliably identified, as their potential combinations involve 5 to 6 parent frequencies, suggesting they may be coincidental.}

\section{DISCUSSION}

\subsection{which star pulsates?}

In asteroseismology, identifying the pulsation modes is a critical step before any seismic inference can be made, as only with correct mode identification can the observed frequencies be reliably compared with those predicted by theoretical models, allowing the internal structure of stars to be accurately determined \citep{2010aste.book.....A}. For a pulsating eclipsing binary, the first step is to determine which component of the system is responsible for the pulsations, as pulsations originating from different components may indicate different evolutionary stages of the system. 

Generally, the temperatures of the component of binary are different. If there are available temperature values for each component, we can preliminarily determine which star the pulsation occurred on, as different classes of pulsating stars often have different ranges of effective temperature. For example, $\delta$ Sct stars occupy a temperature range of 6500 to 8500 K. However, for KIC 5623923, as there is currently no available temperature information on the two components of this system, we used the broad-band spectral energy distribution (SED) fitting to constrain the temperature of this target. During SED fitting, we utilize the BT-Settl \citep{2012RSPTA.370.2765A} synthetic spectral library to predict the bandpass-average magnitudes, while the spectral models are interpolated by package $\tt stellarSpecModel$\footnote{https://github.com/zzxihep/stellarSpecModel}. The MCMC Ensemble sampler $\tt emcee$ \citep{2013PASP..125..306F} is employed to sample different parameters, mainly including temperature, radius, and extinction. We also assume the \citet{1999PASP..111...63F} extinction law with a canonical interstellar reddening law $R_{\rm V} = 3.1$ \citep{1989ApJ...345..245C}. For this target, we adopt the three-dimensional (3D) reddening map from \citet{2019ApJ...887...93G} to acquire the color excess values $E(B-V) = 0.11$ ($A_{\rm V} = 3.1 \times 0.11 = 0.34$ mag). The $\log g$ for both components is assumed to be 4.0 because the SEDs are weakly sensitive to surface gravity \citep{2022MNRAS.515.1266E}. The parallax 0.1055 $\pm$0.040 mas from Gaia DR3 \citep{2023A&A...674A...1G} is treated as the prior of distance. As listed in Table~\ref{tab:obs_mags} sorted from shortest to longest central wavelengths, we collect the multi-band photometric observations from UCAC4 \citep[B and V band;][]{2012yCat.1322....0Z,2013AJ....145...44Z}, Gaia DR3 \citep[G$_{\rm BP}$, G and G$_{\rm RP}$ bands;][]{2022yCat.1355....0G, 2023A&A...674A...1G}, TESS \citep{2019AJ....158..138S}, 2MASS \citep[J, H and Ks bands;][]{2006AJ....131.1163S}. The total system temperature represented by $T_{\rm GTC}$ (Table~\ref{tab:basic_parmeters}) is used as a prior for the SED fitting, with the prior modeled as a normal distribution with $\mu = 7970\,$K and $\sigma = 110\,$K.

\begin{table*}
\begin{center}
\caption{The observational magnitudes of KIC 5623923 in different bands.
\label{tab:obs_mags}}
\begin{tabular}{llccc} \hline \hline
\noalign{\smallskip}
Bands &  $M_{\rm obs}$  &  $\sigma_{\rm M}$ &  Ref. \\
      &  (mag)          &  (mag)              &      \\
\hline
\noalign{\smallskip}
B  & 15.327 & 0.036 & (1) \\  
G$_{\rm BP}$  & 16.585  & 0.004  &  (2)  \\  
V  & 16.501 & 0.024  &  (1)  \\  
R  & 16.504 & 0.024  &  (1)  \\  
r$_{\rm SDSS}$  & 16.473 & 0.009   & (3)  \\  
G  & 16.420 & 0.003  & (2)  \\   
i$_{\rm SDSS}$    & 16.507  & 0.007  &  (3)  \\   
G$_{\rm RP}$   & 16.141 & 0.004  &  (2)  \\  
TESS    & 16.209  & 0.011  & (4) \\   
z$_{\rm SDSS}$    & 16.567  & 0.002   &  (3)  \\  
y$_{\rm PS}$   & 16.485  & 0.006  &   (5)   \\   
J   & 15.869   & 0.073  & (6) \\   
H   & 15.631  & 0.110   & (6)  \\  
Ks  & 15.542  & 0.227   & (6) \\  
Parallaxes (mas) & 0.1055&  0.04 & (2)\\
\noalign{\smallskip}            
\hline   
\end{tabular}
\end{center}
\tablecomments{(1): \citealt{2012yCat.1322....0Z}, UCAC4; (2): \citealt{2022yCat.1355....0G}, Gaia DR3;  (3): \citealt{2015AAS...22533616H}, SDSS; 
  (4): \citealt{2019AJ....158..138S}, TESS; (5): \citealt{2016arXiv161205560C}, Pan-STARRS; (6): \citealt{2006AJ....131.1163S}, 2MASS.  }
\end{table*}

\begin{figure*}
\centering
\includegraphics[width=0.87\textwidth,trim=25 5 0 10,clip]{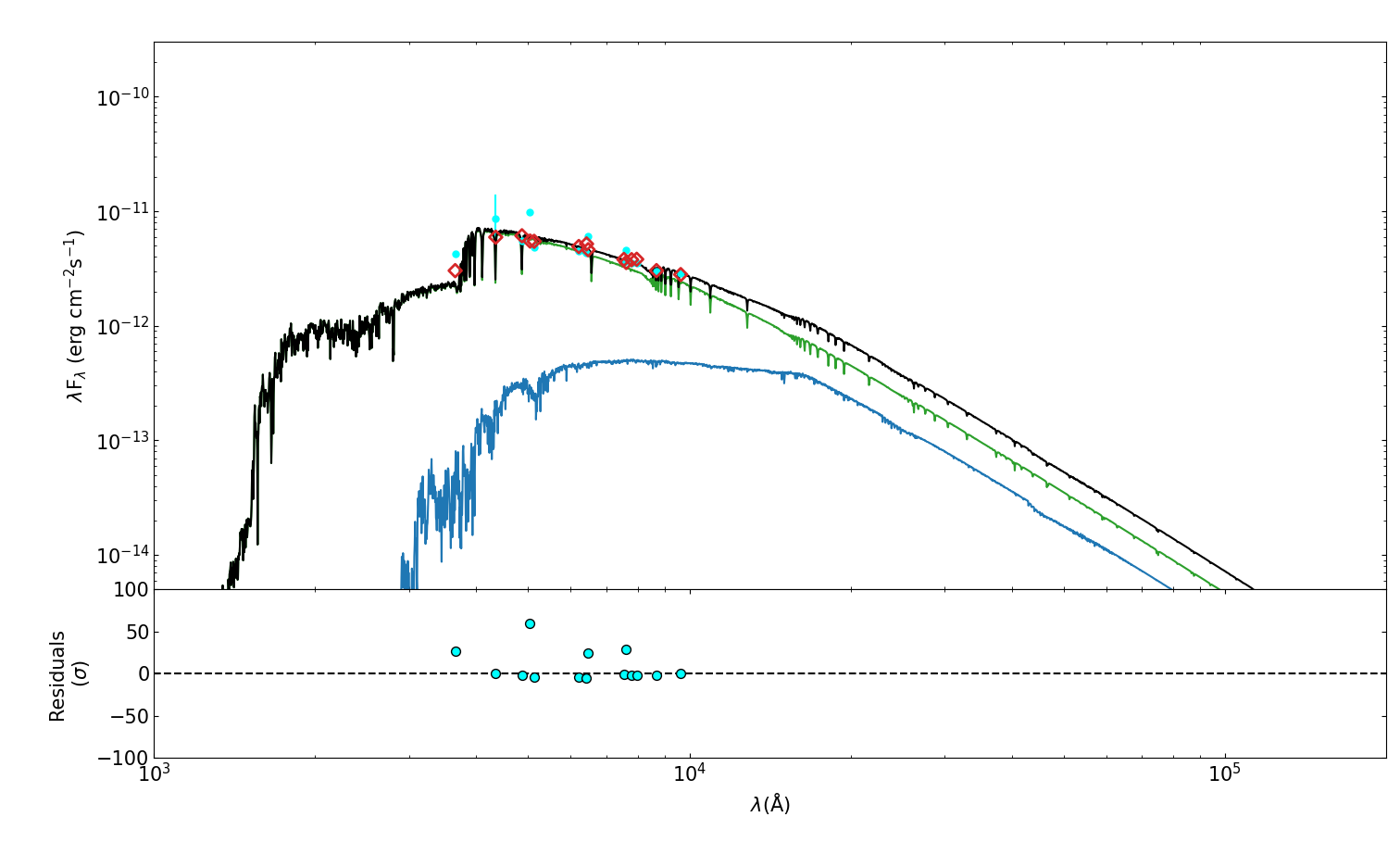}
\caption{The SED fitting result. Upper panel: Cyan points are the observed fluxes, and red diamonds are the synthetic fluxes. The black spectrum is the best-fit SED model, while the green and blue curves are the model spectra of primary and secondary stars, respectively. Lower panel: the residuals of SED fitting. The photometric observations in different bands are sorted in the order of Table~\ref{tab:obs_mags}.}
\label{fig:SED}
\end{figure*}
\begin{figure*}
\centering
\includegraphics[width=0.87\textwidth,trim=5 5 30 0,clip]{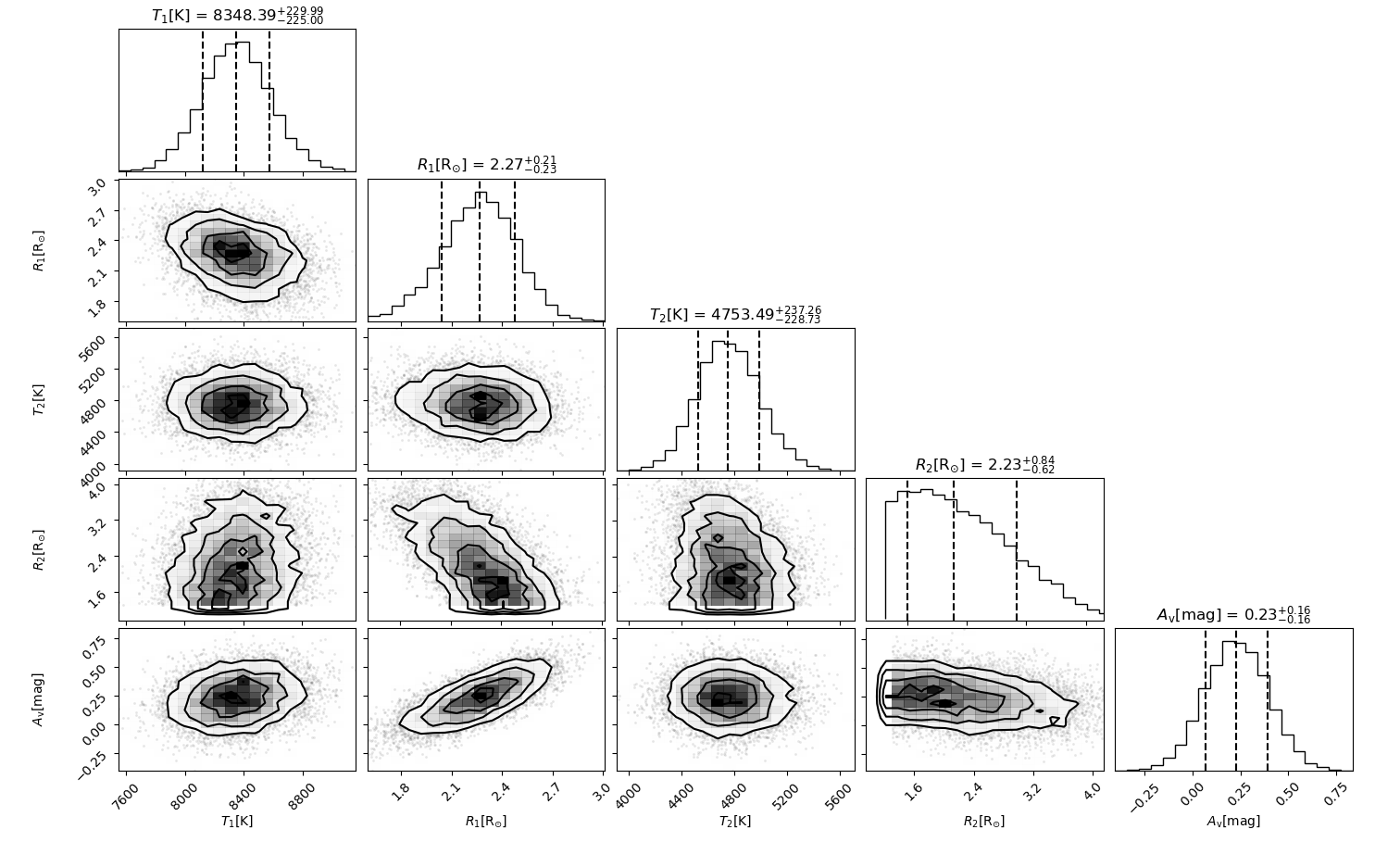}
\caption{The marginalized posterior probability distributions of temperatures, radii and V-band extinction in SED fitting.  The vertical dashed lines are located at the 16th, 50th and 84th percentiles.}
\label{fig:Distribution_T_Av}
\end{figure*}

The SED fitting result is shown in Figure~\ref{fig:SED}, while the marginalized posterior probability distributions of temperatures, radii, and extinction in the V-band are plotted in Figure~\ref{fig:Distribution_T_Av}. When the temperatures of the primary and secondary stars converge to $8348^{+230}_{-225}$~K and $4753^{+237}_{-229}$~K, respectively, the SED model effectively describes the observed photometric data in different bands. Furthermore, the radii are 2.27$^{+0.21}_{-0.23}\rm R_{\odot}$ and 2.23$^{+0.84}_{-0.62}\rm R_{\odot}$, respectively. Based on these best-fitting results, we also obtain the luminosities of the primary and secondary stars as log $L_{1}/L_{\odot}$ = 1.35 $\pm$ 0.09 and log $L_{2}/L_{\odot}$ = 0.36 $\pm$ 0.22, respectively. Therefore, we found that the primary star lies in the $\delta$ Scuti instability strip \citep{2011A&A...534A.125U,2015AJ....149...68B,2017ChA&A..41..471X}; however, the secondary star is far away from the instability strip, indicating that the pulsations originate from the hotter primary star. We note our finding is not accord to the results given by \cite{2014MNRAS.437..132R}.

{Furthermore, we determined the posterior distributions of the system parameters using the neural network (NN) models in tandem with the MCMC approach introduced by \citet{2025AJ....169..202D}. This method employs a neural network trained on a large set of synthetic light curves of semidetached binaries, generated with the PHOEBE code \citep{2016ApJS..227...29P}. The NN has nine input nodes, corresponding to the following astrophysical parameters: the effective temperatures of the primary and secondary stars ($T_1$, $T_2$), orbital inclination ($incl$), mass ratio $q = m_2/m_1$, relative radii of the primary and secondary $r_1 = R_1/a$ and $r_2 = R_2/a$ (where $a$ is the semi-major axis), and four starspot parameters: colatitude ($colat$), longitude ($long$), angular radius ($radius$), and temperature factor ($relT_{\mathrm{eff}}$). }

{A slight asymmetry near the secondary minimum in the light curve suggests the presence of a starspot. We therefore included a single spot in the light curve model. In the fit, the primary effective temperature $T_1$ was fixed at 8348~K, as derived from SED fitting. The free parameters were $T_2$, $incl$, $q$, $r_1$, along with the four spot parameters ($colat$, $long$, $radius$, $relT_{\mathrm{eff}}$). We also allowed for small adjustments in phase shift and magnitude zero-point ($ShiftPhase$, $ShiftMag$).}

\begin{figure*}
\centering
\includegraphics[width=0.87\textwidth,trim=0 0 0 0,clip]{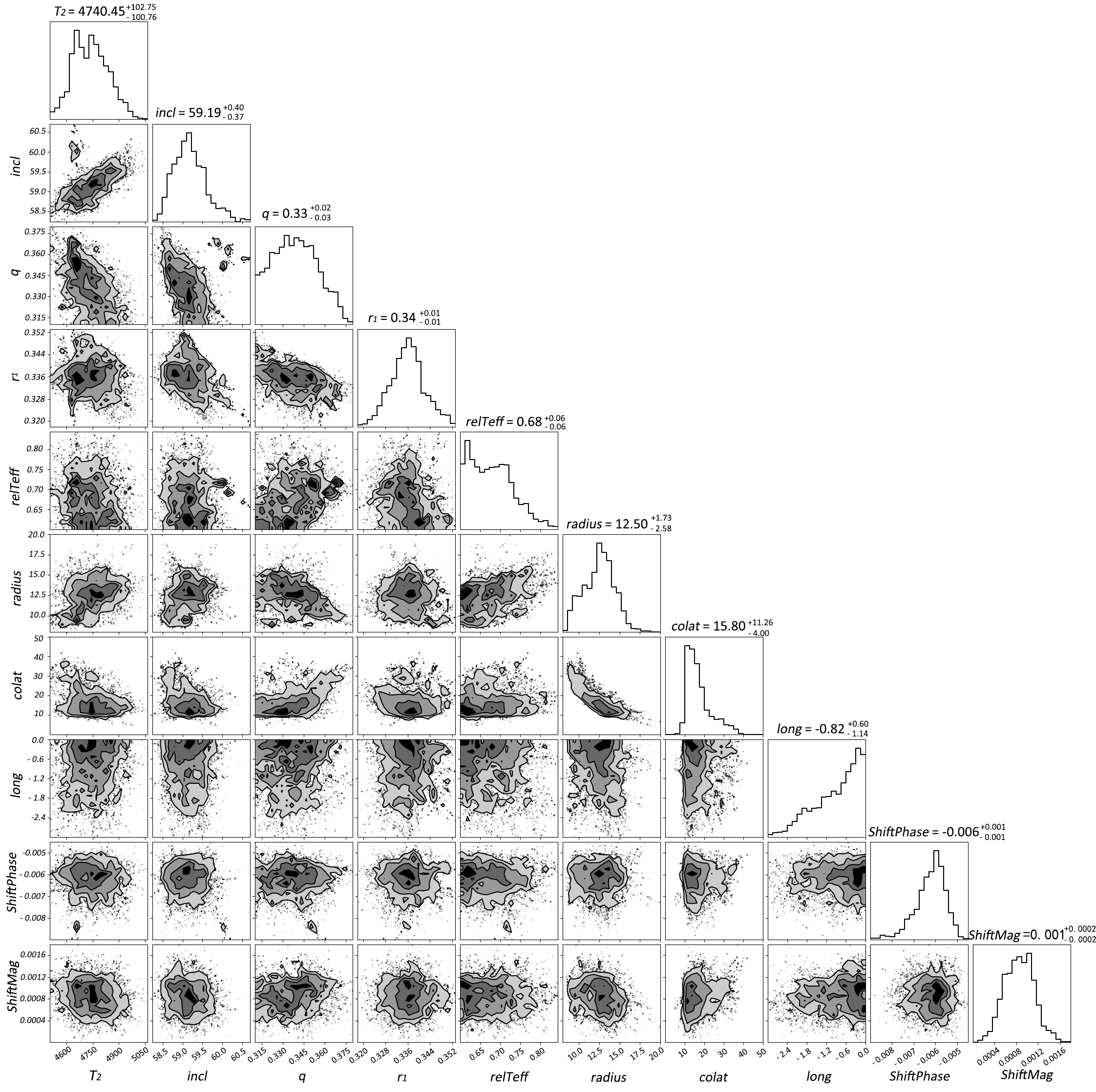}
\caption{The posterior parameter distributions for KIC 5623923 derived by neural network (NN) model.}
\label{fig:AI_phoebe}
\end{figure*}

\begin{figure*}
\centering
\includegraphics[width=0.87\textwidth,trim=0 0 0 0,clip]{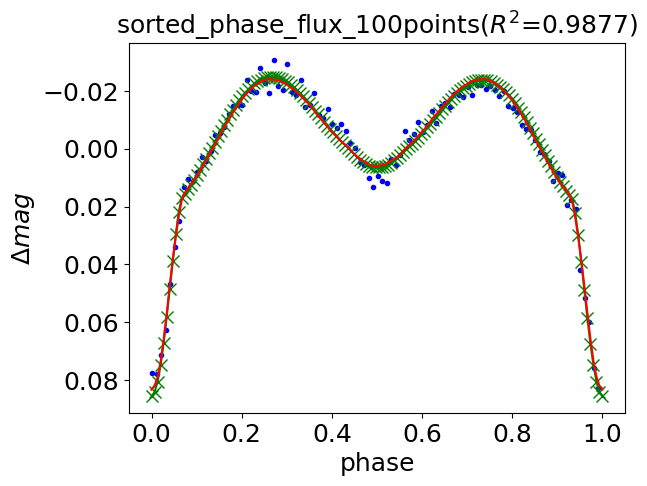}
\caption{Result of light curve fitting for KIC 5623923. Blue dots represent the observed light curves. The green “x” points correspond to the light curve generated by PHOEBE with identical parameters. The red line represents the light curve reconstructed by NN model using the derived parameters. The goodness of fit ($R^2$) between the PHOEBE-generated light curve and the observed light curve was found to be 0.9877. }
\label{fig:AI_lc}
\end{figure*}

{Figure~\ref{fig:AI_phoebe} shows the resulting posterior distributions of the fitted parameters, while Figure~\ref{fig:AI_lc} compares the observed light curve (blue points) with the NN-reconstructed model (red line). The fitted temperature of the secondary, $T_2 = 4740 ^{+102.75}_{-100.76}$~K, is consistent with the SED result ($4753.49 ^{+237.26}_{-228.73}$~K). However, in the absence of radial velocity measurements to constrain the semi-major axis $a$, we report only the relative radius of the primary, $r_1 = 0.34$.}

\begin{figure}
  \centering
  \includegraphics[width=0.95\textwidth,trim=30 245 25 245,clip]{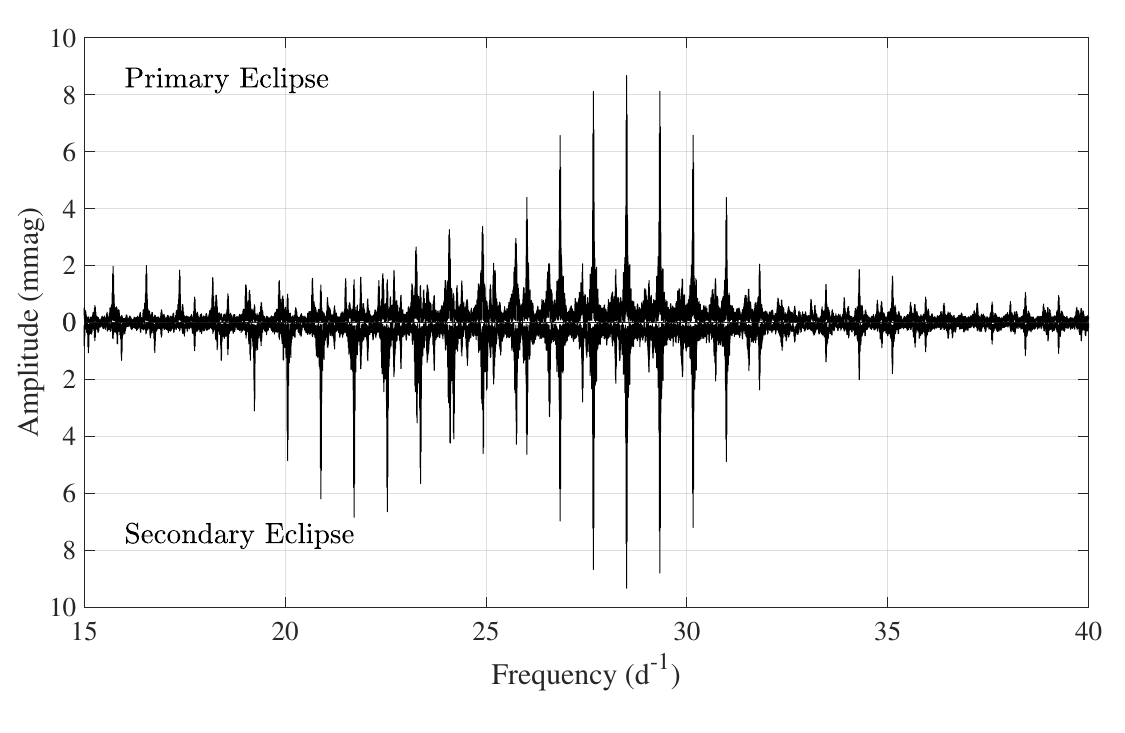}
  \caption{Amplitude spectra of the light curve during the times of the eclipses. The upper panel shows the amplitude spectrum during primary eclipse, and the inverted lower panel shows the amplitude spectrum during the secondary eclipse.}
   \label{fig:frequency_spectra_two_panel}
\end{figure}

In order to discern what the contribution of each star to the observed frequency spectrum is, we examined the light curve at the times of the eclipses. We extract two light curves around each primary and secondary minima in a time length of 0.2 periods, respectively, and perform the Fourier transform for them, as shown in Figure \ref{fig:frequency_spectra_two_panel}. From this figure, it clearly shows that the amplitude around the primary minima is weaker than that around the secondary minima. This is because, during the primary eclipse, the secondary component partially or completely obscures the light from the $\delta$ Scuti star, leading to a weakening effect on the observed amplitudes of its pulsation modes due to geometric occlusion. This provides additional evidence that the $\delta$ Scuti pulsations originate from the hotter primary star.     

\subsection{Multiplet structure.}

Recently, multiplet structures in frequency spectra have been observed in various types of pulsating variables, thanks to the high-precision and nearly continuous time-series photometry provided by space telescopes such as $Kepler$ and $TESS$. These multiplet structures offer additional constraints on the stellar parameters. For example, in the HADS star KIC 5950759, a pair of equally spaced triplet structures centered on the main frequency were detected in its frequency spectrum. The triplet structure is interpreted as the result of amplitude modulation due to stellar rotation at 0.3193 d$^{-1}$, thereby providing a means to constrain the star’s rotational properties \citep{2018ApJ...863..195Y}. In addition, another HADS star, KIC 10284901 also shows the quintuplet structures around the main pulsation modes, and analysis suggests that they might be related to the Blazhko effect \citep{2019ApJ...879...59Y}. However, such multiplet structures have typically been observed in single stars. In binary systems, particularly eclipsing binaries, thanks to the high-precision and long time-span photometric data provided by $Kepler$, the effect of orbital motion on stellar pulsations can be measured, leading to two new methods, the frequency modulation (FM) method \citep{2012MNRAS.422..738S,2015MNRAS.446.1223K,2015MNRAS.450.3999S} and phase modulation (PM) method \citep{2014MNRAS.441.2515M,2014MNRAS.443.1946B,2014MNRAS.444.1486K}.      

In the case of KIC 5623923, we attempted to investigate whether the pulsation frequencies exhibit a detectable time-delay curve. {However, both the frequency modulation (FM) and phase modulation (PM) methods proved ineffective due to the system's extremely short orbital period ($P_{\mathrm{orb}} = 1.2089$~d). Assuming that the system has a circular orbit ($e=0$), the primary star is a typical $\delta$ Sct star with mass of 1.8 M$_{\odot}$, and adopting the NN-derived parameters -- mass ratio $q=0.33$, inclination $i=60^\circ$, and relative radius $r_1 = 0.34$ -- we calculate a semi-major axis $a \approx 6.39~R_\odot$ and primary semi-major axis $a_1 = 1.59~R_\odot$. This yields a maximum light travel time delay of $a_1 \sin i/c \approx 3.18$~s, which falls below the $\sim$10~s detection limit established by \citet{2016MNRAS.461.4215M} for 1500-day photometric data. Given that our observational baseline spans only $\sim$307 days, the current data are insufficient to detect measurable time-delay effects, even for radial pulsation modes.} We suggest that future space-based observations (e.g. ET2.0: \citet{2024ChJSS..44..400G}; PLATO: \citet{2024arXiv240605447R}) of this target, with longer time baselines and higher frequency resolution, will be essential to confirm this result and to investigate the potential influence of orbital motion with greater certainty.

\subsection{Mode identification.}


{Tidal synchronization is expected in close binary systems due to orbital evolution under tidal forces \citep{1981A&A....99..126H,2008CeMDA.101..171F,2010A&A...516A..64L,2021AJ....161..263Z}. Statistical studies support this expectation: \cite{2017AJ....154..250L} analyzed 2278 Kepler eclipsing binaries and found that most systems with orbital periods below 10 days are tidally synchronized. Given its short orbital period ($P_{\mathrm{orb}} = 1.2089$ days), KIC 5623923 is therefore highly likely to be tidally synchronized. Further evidence comes from the light curve morphology: the clear ellipsoidal variations visible in Figure \ref{fig:SC_phase} indicate tidal distortion of the components, with the characteristic double-peaked pattern occurring at orbital quadratures when the stars present their largest cross-sectional areas \citep{2025ApJ...990..124H}. This combination of short period and photometric signature strongly supports the case for tidal synchronization in this system.}

In asteroseismology, mode identification is a crucial yet challenging step, particularly for $\delta$ Scuti stars, which typically exhibit rich and complex amplitude spectra. Eclipsing binary systems that contain $\delta$ Scuti-type pulsators provide valuable opportunities for reliable mode identification. To aid this process, \citet{2005ApJ...634..602R} developed a tool specifically designed for the asteroseismic analysis of pulsating stars in close binary systems. In their work, assuming that spherical harmonics adequately describe the surface geometry and that stellar rotation is synchronized with the binary orbit, as in the case of KIC 5623923, they investigated how variations in the inclination angles of the pulsation geometry and the rotation axis affect the observed pulsation modes from an observational standpoint.

In their study, one of the key findings is that radial modes are completely unaffected by either the inclination of the pulsation axis (as they actually have none) or by eclipses. Specifically, radial modes do not exhibit amplitude variations with orientation or orbital phase and, therefore do not produce multiplet structures in the frequency spectrum (the reason has been discussed in Section 4.2). Consequently, in the case of KIC 5623923, the strongest peak, $f_{3}$, can be considered to be a radial mode. {Using the $\delta$~Sct period--luminosity relation from \citet{2019MNRAS.486.4348Z} and the derived primary luminosity $\log(L_1/L_\odot) = 1.35 \pm 0.09$, the predicted frequencies for the fundamental, first, second, and third overtone radial modes are $f_1 = 8.49 \pm 1.59$~d$^{-1}$, $f_2 = 10.99 \pm 2.06$~d$^{-1}$, $f_3 = 13.69 \pm 2.57$~d$^{-1}$, and $f_4 = 16.59 \pm 3.11$~d$^{-1}$, respectively. All of these are significantly lower than the observed strongest frequency ($28.499399$~d$^{-1}$), indicating that it may belong to a high-order mode.}

However, for non-radial modes, such as those with $l = 1$ or $l = 2$, their frequency spectra can exhibit triplet or multiplet structures with frequency separations equal to the orbital frequency under specific conditions (cf. Figs. 7 and 8 in \citealt{2005ApJ...634..602R}). For example, in the special case where the pulsation axis has an inclination of $I_{p} = 90^\circ$ or the rotation axis has $I_{r} = 90^\circ$, modes with $(l, m) = (1, 0)$ would completely lack the central frequency peak, leaving only orbital aliases. In other scenarios, such as when $I_{p} = 45^\circ$ and $I_{r} = 60^\circ$, triplet structures with unequal amplitudes can appear, as illustrated in Fig. 7 of \citet{2005ApJ...634..602R}. For modes with $l = 2$, the situation becomes more complex. The frequency spectra can exhibit quintuplet structures with varying amplitudes, depending on the values of the pulsation axis inclination ($I_{p}$) and the rotation axis inclination ($I_{r}$). For instance, when $I_{r} = 75^\circ$, the strongest peak in the quintuplet corresponds to the actual pulsation mode with $l = 2$, while the remaining peaks arise from orbital aliases (cf. the second column of Fig. 8 in \citealt{2005ApJ...634..602R}). However, when $I_{r} = 60^\circ$, the intrinsic pulsation mode becomes the weakest peak, and the orbital aliases dominate with higher amplitudes (cf. the third column of Fig. 8 in \citealt{2005ApJ...634..602R}).

{Therefore, based on the observational characteristics of the frequency spectra of KIC 5623923—specifically the quintuplet structures with frequency spacings of $f_{\mathrm{orb}}$ and varying amplitudes, as shown in Figure \ref{fig:SC_spectra2}—it is reasonable to identify $f_{7}$, $f_{10}$, and $f_{18}$ as non-radial modes with $l = 2$. These frequencies correspond to the central components of the quintuplet-like structures, while the associated sidelobes are attributed to orbital modulation, as explained by \citet{2005ApJ...634..602R}.} Considering that this system has undergone mass transfer, the internal structure of the pulsating primary star may have been significantly altered, complicating its asteroseismic modeling. Therefore, a detailed asteroseismic analysis of the primary star is worthy to be investigated further in the future, however is beyond the scope of this study.

\section{CONCLUSIONS}

In this paper, we investigate the light variation of KIC 5623923 using the $Kepler$ short-cadence time series data obtained from Quarters 14 to 17. With Fourier transformation of light curves, we extracted 41 significant frequencies in total, including the orbital frequency ($f_{orb}$ = 0.827198 d$^{-1}$, harmonics up to 18$f_{orb}$) and multiple pulsational frequencies. Most of the pulsational frequencies lie in the frequency range of 20.0 $-$ 35.0 d$^{-1}$, seven of which are identified as independent p modes.   

In the frequency spectrum, we found three groups of frequencies that form quintet-like structures, centered on $f_{7}$, $f_{10}$, and $f_{18}$, but the strongest frequency ($f_{3}$) does not. {Based on the short orbital period and the observed light curve morphology, we conclude that KIC 5623923 is a tidally synchronized system.} Using the characteristics of the amplitude spectral of pulsating stars in close binaries presented by \citet{2005ApJ...634..602R}, we compared that with the multiplet structures detected in this target, and found that three frequencies (i.e. $f_{7}$, $f_{10}$, and $f_{18}$) should belong to non-radial mode with l=2, and $f_{3}$ {may be a high-order} radial mode.

Using collected multi–passband photometry from the literature, we performed a SED fitting for KIC 5623923 and derived effective temperatures of $T_{\rm eff,1}=8348^{+230}_{-225} ~$K and $T_{\rm eff,2}=4753^{+237}_{-229} ~$K for the primary and secondary components, respectively. {Additionally, we employed neural network (NN) models to fit the light curves, yielding system parameters that are consistent with the SED fitting results.} Together with the distance from Gaia DR3, the luminosities of the two components are estimated to be log ($L_{1}/L_{\odot}$) = 1.35 $\pm$ 0.09 and log ($L_{2}/L_{\odot}$) = 0.36 $\pm$ 0.22, respectively. These values place the primary star within the classical pulsating instability strip. Therefore, this system is suggested to be an eclipsing binary with $\delta$ Scuti-type pulsations that originate from the primary component, rather than from the secondary component as previously reported.  

Regarding future studies of KIC 5623923, it is worth noting that radial velocity measurements will help constrain the absolute parameters of the system, such as the masses and surface gravities of the components. Given the faintness of the system ($V = 16.5$ mag), high-resolution spectrographs mounted on 8–10 meter class telescopes are required to detect the spectral signatures of both components.

With the launch of upcoming space survey missions such as the Chinese Space Station Telescope (CSST, \citealt{2018MNRAS.480.2178C}) and PLATO \citep{2024arXiv240605447R}, an increasing number of faint stars are expected to be discovered. By applying the method used in this study, a systematic search for pulsations in faint binary systems, followed by asteroseismic analysis, will provide valuable insights into the formation and evolutionary scenarios of binary stars.

\section*{Acknowledgments}

We are grateful to the referee for constructive comments that improved this manuscript. This research is supported by the National Natural Science Foundation of China (grant Nos. 12003020, 12473043) and Shaanxi Fundamental Science Research Project for Mathematics and Physics (Grant No. 23JSY015), and the National Key R$\&$D Program of China (2023YFB3906000), the National Key R$\&$D program of China for the Intergovernmental Scientific and Technological Innovation Cooperation Project under No. 2022YFE0126200, and the Tian-shan Talent Training Program under No. 2023TSYCLJ0053. We thank Simon Murphy for his valuable insights regarding the application of frequency and phase modulation methods for detecting time-delay curves.

\bibliography{reference}{}
\bibliographystyle{aasjournal}


\end{document}